\documentclass[
  a4paper,
  fontsize=11pt,
  captions=tableheading,
  parskip=never,
  ]{scrartcl}

\setkomafont{caption}{\small}
\setkomafont{captionlabel}{\bfseries}
\addtokomafont{publishers}{\large}
\addtokomafont{subject}{\mdseries\large}

\pdfoutput=1

\usepackage{ILD}
\usepackage{graphicx}

\usepackage[utf8]{inputenx}
\usepackage[T1]{fontenc}
\usepackage[british]{babel}
\usepackage{csquotes}

\usepackage{subcaption}
\usepackage{import}
\usepackage{xspace}
\usepackage{graphicx}
\usepackage[shortcuts]{extdash}
\usepackage{siunitx}
\DeclareSIUnit \lightspeed {\text{{c}}}
\usepackage{xcolor}
\definecolor{linkblue}{HTML}{264772}
\usepackage{hyperref}
\hypersetup{
    colorlinks=true,
    linktocpage=true,
    linkcolor=linkblue,
    citecolor=linkblue,
    urlcolor=linkblue
}

\usepackage{wrapfig}

\graphicspath{ {./Pictures/} }
\DeclareGraphicsExtensions{.pdf,.png,.jpg}

\begin{document}

\hyphenation{
  am-pli-fi-ca-tion
  col-lab-o-ra-tion
  per-for-mance
  sat-u-rat-ed
  se-lect-ed
  spec-i-fied
}

\title{Searching for new physics in WW and single-W events}
\author{Jenny List, Ulrich Einhaus, Andre Filipe Silva, Leonhard Reichenbach}

\titlecomment{Talk presented at the International Workshop on Future Linear Colliders (LCWS 2024), 8-11 July 2024, Tokyo. This work was carried out in the framework of the ILD Concept Group.}
\date{}

\addauthor{Jenny List}{\institute{1}}
\addauthor{Ulrich Einhaus}{\institute{1}\institute{2}}
\addauthor{Andre Filipe Silva}{\institute{1}\institute{3}}
\addauthor{Leonhard Reichenbach}{\institute{4}\institute{5}}
\addinstitute{1}{Deutsches Elektronen\-/Synchrotron DESY, Germany}
\addinstitute{2}{now at: Karlsruhe Institut für Technologie KIT}
\addinstitute{3}{University of Coimbra}
\addinstitute{4}{University of Bonn}
\addinstitute{5}{CERN}

\abstract{

Pair-production and single-production of $W$ bosons provide many opportunities to look for new physics via precision measurements, for instance via scrutinising the involved triple-gauge vertices or by measuring CKM matrix elements in an environment very complementary to $B$ hadron decays. This contribution presents the ongoing work based on full simulation of the ILD concept, exploiting the O($10^8$) $W$ bosons produced during the $250$\,GeV stage of the ILC, as well as the CLD concept proposed for the FCC-ee. The projections, which also contribute to two focus topics of the current ECFA study on future Higgs/Top/Electroweak factories, promise improvements of the measurement precision of TGCs and the CKM matrix, respectively, of one to two orders of magnitude with respects to LEP.}

\titlepage
\clearpage

\section{Introduction}
The international particle physics community agrees that the highest priority next collider should be an $e^+e^-$ Higgs factory. At such a collider also $W$ bosons would be produced copiously: for instance the ILC would produce $1.2 \times 10^8$ $W$ bosons at $250$ and $500$\,GeV, while the FCC-ee program offers similar amounts at $160$ and $240$\,GeV. The production processes comprise $W$ pair production and single-$W$ production, i.e.\ the t-channel contribution with the tree-level Feynman diagram containing only one $W$, hence called the single-$W$ channel. Both processes are highly sensitive to the beam polarisation, and both suitable to scrutinize triple-gauge vertices connecting two $W$ bosons with a $Z$ boson or a photon. In addition, the hadronic decays of the $W$ bosons probe elements of the CKM matrix -- a quite complementary approach to $B$ factory measurements, in particular concerning the involved systematic uncertainties.
%In recent years, the focus has shifted from linear to circular machines, and thus from highest energies to highest luminosities.
%This has been accompanied by an increased interest in more detailed studies of the  electroweak sector in addition to the more established ones of the Higgs and top, hence the name Higgs/top/electroweak (HTE) factory.
%Such electroweak studies entail $Z$ and $W$ physics, both with several orders of magnitude more statistics compared to LEP, specifically some $10^8$ $W$s with both a linear or circular $e^+e^-$ collider.

%The currently ongoing ECFA study on a future HTE factory has defined 14 focus topics \cite{ECFAFocusTopics}, three of which target the $W$: the $W$ mass, the differential production and decay cross sections of the $W$, in particular involving triple gauge couplings (TGCs) involving the $W$, and a determination of the 6 CKM matrix elements that do not involve the top, directly from hadronic $W$ decays.
This contribution gives an overview of ongoing work and highlights the prospects for their measurements at a future $e^+e^-$ collider.

\section{Triple Gauge Couplings}
In its most general form, the vertex between a neutral and two charged electroweak gauge bosons includes 14 complex couplings, i.e.\ 28 real parameters. In the Standard Model (SM), only four of them, $g_{1,Z}$, $g_{1,\gamma}$, $\kappa_Z$ and $\kappa_{\gamma}$, are equal to one, while all others are predicted to be zero. Any deviation from this prediction presents a highly sensitive probe for physics beyond the SM. Since the LEP data were not sufficient to constrain the full set of TGCs, the number of parameters usually considered is strongly reduced by demanding $C$, $P$ and $CP$ invariance, as well as electromagnetic gauge and $SU(2)\times U(1)$ invariance, leaving three free parameters, e.g.\ $g_{1,Z}$, $\kappa_{\gamma}$ and $\lambda_{\gamma}$ as used at LEP, or the equivalent parameters in SMEFT. $W$ pair production in $e^+e^-$ collisions offers and ideal laboratory: in addition to the total polarised cross-sections, the differential cross-section in terms of the production angle as well as the two times two decay angles of the $W$ bosons give sensitivity to the triple-gauge vertices. Any projection will depend significantly on the exact analysis technique applied: whether the analysis of the differential distributions is performed binned or un-binned, whether three or all five angles are used, whether single- and multi-parameter fits are done, and how which systematic uncertainties are included.

A number of studies on triple-gauge couplings at future colliders have been performed, from theory-level fits to optimal observables to studies in full, Geant4-based, detector simulations. The most comprehensive study, purely on theory-level, i.e.\ considering neither detector resolution, nor backgrounds, nor systematic uncertainties, showed that at a high-energy $e^+e^-$ collider at $\sqrt{s}=500$\,GeV with polarised beams, including transverse polarisation, in principle all 28 real coupling parameters can be disentangled and constrained~\cite{Diehl:2002nj, Diehl:2003qz}. More recently, the sensitivity to the three couplings of the LEP parametrisation has been studied in the context of a SMEFT fit  to the Higgs and electroweak sector, also based on theory-level optimal observables~\cite{deBlas:2022ofj}. This study projects an impressive improvement of the precision by about a factor of more than a $100$ compared to HL-LHC. In particular for $\lambda_{\gamma}$, the sensitivity does improve significantly with energy, so that the TeV-scale energy stages of linear colliders even improve by about a factor of $1000$ over HL-LHC. The effect of detector resolution and residual backgrounds on the optimal observable technique has been studied in~\cite{Chai:2024zyl}, showing that these effects -- if uncorrected for -- lead to significant biases of the central values obtained from the optimal observable technique, and demonstrating that for instance machine-learning techniques can be employed to predict the necessary corrections.

Studies based on realistic detector simulations only exist for ILC at $\sqrt{s}=500$\,GeV and $1$\,TeV~\cite{Marchesini:2011aka, ILC_TDR_Detectors}. Due to the limited available capacity for Geant4-based Monte-Carlo production at the time, only $W$ pair production and three out of the five angular observables were considered in a binned analysis, rendering the results extremely conservative. Nevertheless, the covariance matrix from the simultaneous fit of the three LEP couplings from these studies was the basis for the input to the Higgs and electroweak fits performed e.g.\ in~\cite{Barklow:2017suo}. Projections for lower center-of-mass energies up to now rely on interpolations between the $500$\,GeV / $1$\,TeV projections and the actual LEP2 results~\cite{Karl:2019hes}. These showed that at $250$\,GeV, the sensitivity is expected to be a factor of 4-5 worse than at $500$\,GeV for the same final-states considered. This study also for the first time highlighted the significant additional gain from including the single-$W$ final states into, improving the precision of the $e\nu qq$ channel by a factor 2-3 compared to including only $WW \rightarrow e\nu qq$ events. A first comprehensive study of experimental uncertainties and how their impact on the final result can be minimised has been presented in~\cite{Beyer:2022ofv}. This work showed when combining data from various 2-fermion and 4-fermion final-states, all involved physics observables (like the TGCs but also e.g.\ forward-backward asymmetries etc.) can be disentangled from systematic effects from e.g.\ luminosity, polarisation, energy, detector acceptance and others in a combined fit with physics and nuisance parameters, with very small residual impact from the systematic uncertainties on the physics observable. This observation holds in particular when both beams are polarised, providing a sufficient number of over-constraining data-sets.

\section{CKM matrix determination from hadronic $W$ decays}

There is at the moment an intriguing discrepancy at the $3\sigma$ level in the value of $|V_{cb}|$ from $B$ decays, being at $(42.19 \pm 0.78) \times 10^{-3}$ from inclusive $B$ decays and at $(39.10 \pm 0.50) \times 10^{-3}$ from exclusive $B$, $B_s$ and $\Lambda_b$ decays.
While this discrepancy is difficult to solve in $B$ decays due to inherent hadronic uncertainties, these are absent in decays of real $W$s.
With the large QCD background, the LHC prospects of this measurement are only at the level of 10$\%$. However, a future Higgs/Top/Electroweak factory would offer a clean environment with $O(10^8)$ $W$ bosons, leading to a ideal uncertainty on the measurement of $|V_{cb}|$ of 0.15$\%$ given 100$\%$ efficiency and no backgrounds.
This is accompanied by theory uncertainties expected to drop to the level of $10^{-4}$ until such measurement is conducted.
Moreover, $|V_{cb}|$ is not the only point of interest, rather all 6 CKM matrix elements that do not involve the top quark can be determined via hadronic $W$ decays in  a manner complementary to $B$ and $D$ meson decays at corresponding factories.
\autoref{tab:CKMprecision} summarises the branching ratios, expected number of $W$ decays for $10^8$ $W$s and the following ideal determination uncertainty for these 6 elements.
While these ideal uncertainties give us an interesting lower limit, detailed physics and detector simulations involving efficiencies and background are needed to determine how close we can get to the ideal numbers.

\begin{table}[h!]
\begin{center}
\begin{tabular}{c|c c c c c c }
 $W^-$ & $\bar{u}d$ & $\bar{u}d$ & $\bar{u}b$ & $\bar{c}d$ & $\bar{c}d$ & $\bar{c}b$\\\hline
 BR & 31.8$\%$ & 1.7$\%$ & $4.5 \times 10^{-6}$ & 1.7$\%$ & 31.7$\%$ & $5.9 \times 10^{-4}$ \\
 $N_{ev}$ & $64 \times 10^6$ & $3.4 \times 10^6$ & 900 & $3.4 \times 10^6$ & $63 \times 10^6$ & $118 \times 10^3$ \\
 $\delta^{th}_{V_{ij}}$ & 0.0063$\%$ & 0.027$\%$ & 1.7$\%$ & 0.027$\%$ & 0.0063$\%$ & 0.15$\%$
\end{tabular}
\caption{$W$ decay branching ratios in the Standard Model, corresponding occurrence for $10^8$ $W$s and CKM matrix element determination precision given 100$\%$ efficiency and no backgrounds; from \cite{ECFAFocusTopics}.}
 \label{tab:CKMprecision}
\end{center}
\end{table}

A first study~\cite{Marzocca:2024dsz} considering 2-fermion processes as background, a parametrised flavour tagging inspired by the IDEA detector concept for FCC-ee and potentially limiting systematic effects showed that for $|V_{cb}|$ a control of the flavour tagging efficiency at the level of about $0.1\%$ would be sufficient, while $|V_{cs}|$ the efficiency would need to be controlled to better than $0.01\%$ in order not to limit the achievable precision.

A more comprehensive set of background processes and systematic uncertainties has been considered in a full-simulation study based on the ILD version for CEPC~\cite{Liang:2024hox}. This study showed that it is not sufficient to consider only 2-fermion backgrounds, since also 4-fermion processes contribute significantly to the final selection. It also showed that systematic uncertainties should improve by about a factor 4 over LEP systematics in order not to limit the final result and that in terms of statistical precision, $5$\,ab$^{-1}$ of unpolarised data give about the same result as $2$\,ab$^{-1}$ of polarised data. 

\section{Ongoing work in ILD/CLD}
ILD and CLD are working towards a full analysis of all $WW$ and single-$W$ channels. A first tool has been developed providing a unique categorisation of each event into the possible $WW$ and single-$W$ decay channels, hadronic, semileptonic and leptonic.
This allows for any $W$-related study to be performed on overlap-free exclusive-channel samples, which are coherent between studies, in order for an easy comparison and combination of their results.

The categorisation algorithm uses the number of isolated leptons in each event, as reconstructed in the ILD~\cite{ILD_IDR} full-simulation and reconstruction chain via isolated lepton tagging algorithms (for electrons and muons) and a dedicated $\tau$ finder.
This also allows for for a sub-categorisation of the semileptonic channel into the three lepton flavours.
Furthermore, a cut on the event invariant mass is used to tag so-called invisible semileptonic events, where the electron from the leptonically decaying $W$ is very forward and outside of the detector acceptance.

\begin{figure}[!htp]
  \centering
  \includegraphics[width=.9\textwidth,keepaspectratio=true]{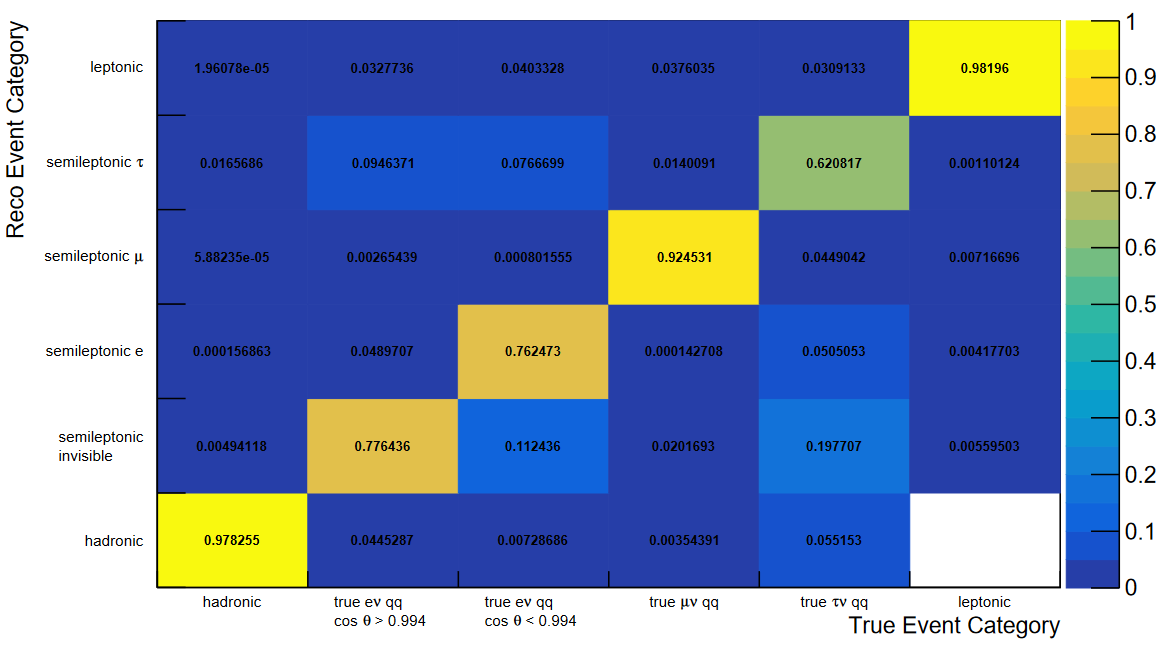}
  \caption{Confusion matrix of the $WW$ categorisation.}
  \label{fig:WWCat_CF}
\end{figure}

This basic categorisation already delivers decent results, but can be easily improved by employing cuts on certain event properties, namely number of reconstructed particles, invariant mass, missing energy and missing transverse momentum, in order to re-categorise events.
The result is shown in \autoref{fig:WWCat_CF}, with the numbers on the diagonal indicating the efficiencies of the category reconstruction.
On first view, the dominant diagonality shows a good overall categorisation.
The hadronic and leptonic channels are reconstructed with very high efficiency of 98$\%$ and the semileptonic muonic channel with 92$\%$.
The semileptonic channels with an electron are categorised moderately well at around 77$\%$, with both channels where the simulated electron is inside or outside or the detector acceptance (cos$\theta < 0.994$) having some confusion with each other, but also with other channels, in particular the semileptonic one with tauons.
This channel itself also only has a moderate efficiency at 62$\%$, 'losing' mostly to the semileptonic-invisible channel.
For a future refinement of this matrix, these points of confusion can serve as a starting point, in particular the $\tau$ finder.

\begin{figure}[!htp]
  \centering
  \includegraphics[width=\textwidth,keepaspectratio=true]{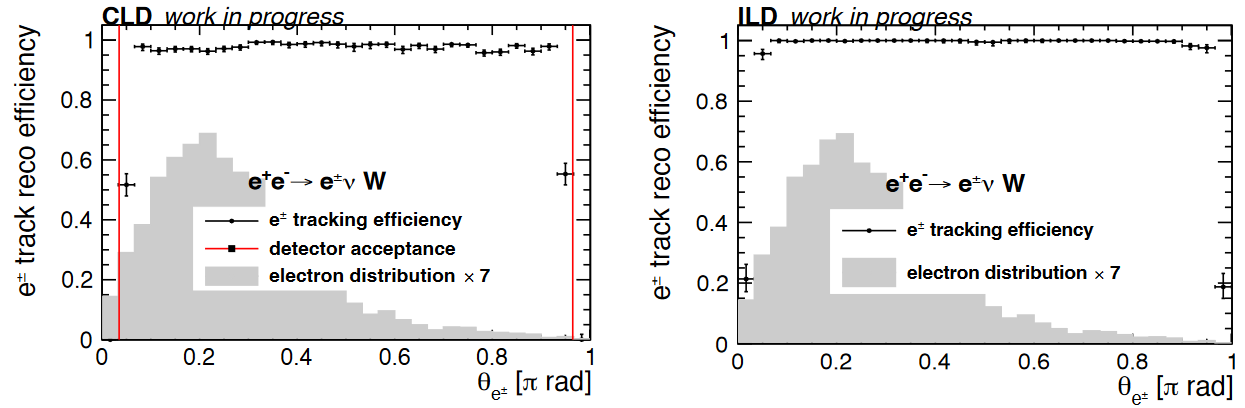}
  \caption{Electron angular tracking efficiency, assessed with the same performance evaluation for CLD and ILD.}
  \label{fig:ForwardTracking}
\end{figure}

The confusion between the visible and invisible semileptonic electron channels play an important role in particular in the $W$ differential cross section measurement, since the angular cut-off in reconstruction is in the region of the largest contribution of the single-$W$ channel.
Dedicated studies of forward-electron reconstruction are being performed, in order to alleviate this confusion and, ideally, improve the overall reconstruction efficiency. This is also shown in \autoref{fig:ForwardTracking}, where in a first step, the same performance evaluation in Key4hep \cite{key4HEP} has been applied to CLD and ILD full-simulation data.
One crucial issue is misreconstruction due to bremsstrahlung, which may be reduced through Gaussian sum filters. The goal here is to implement these in ACTS (A Common Tracking Software \cite{ACTS, ACTS_sw}) in key4hep and then apply the same tracking framework to all Higgs factory detector concepts that provide full detector simulation.

The measurement of CKM matrix elements from hadronic $W$ decays requires excellent flavour tagging, including dedicated strange tagging.
Several new flavour taggers which are all utilising modern machine learning techniques are under development and can address this question.
One of these \cite{FlavourTagger_Mareike}, based on ParticleNet, is being adapted to allow for strange tagging.
This is enabled by utilising particle identification (PID), in particular kaon ID, since jets from strange quarks have a higher average fraction of strange hadrons, in particular after weighting with momentum.
The PID information is provided by the new tool CPID \cite{CPID} and both this and the inference part of the flavour tagger have been recently implemented in the ILD reconstruction chain.

\section{Conclusion}
This contribution gave an overview of existing results and ongoing work quantifying the capabilities of future $e^+e^-$ colliders in the areas of constraining triple gauge couplings and CKM matrix elements from $W$ boson production. In both areas, very significant progress is expected beyond today's state-of-the-art, as well as with regard to expectations from HL-LHC and $B$ factories. For some of the observables, the ability to measure at the higher center-of-mass energies offered only by linear colliders gives a distinct advantage. The measurements require the clean environment of an $e^+e^-$ collider and are not expected to be superseeded by a future hadron collider.

%\newpage
\bibliographystyle{abbrv_mod}
\bibliography{References}
%\printbibliography
\end{document}